\documentclass[twocolumn,superscriptaddress,floatfix,preprintnumbers]{revtex4-1}
\usepackage{times}
\usepackage{mathptmx}
\usepackage{graphicx}
\usepackage{xspace}
\usepackage{amsmath}
\usepackage{pgffor} % To support \foreach coommand in macros

\usepackage{tabularx}
\newcolumntype{C}[1]{>{\centering}p{#1}}
\newcolumntype{L}[1]{>{\raggedright}p{#1}}
\newcolumntype{R}[1]{>{\raggedleft}p{#1}}

\usepackage{MnSymbol}
\usepackage{pifont}

\newcommand{\papertitle}{Phase-channel dynamics reveal the role of impurities and screening in a quasi-one-dimensional charge-density wave system} 

\newcommand{\ffm}{Physikalisches Institut, J. W. Goethe-Universit\"at, 60438 Frankfurt am Main, Germany} 

\newcommand{\usr}[3]{\ensuremath{#1_{\mathrm{#2}}^{\mathrm{#3}}}} % helper for upright subscripts/superscripts

\newcommand{\unit}[1]{\ensuremath{\;\mathrm{#1}}}

\DeclareRobustCommand{\units}[1]{%
\ensuremath{\, \foreach \x/\y in {#1} {{\,\mathrm{\x}^{\y}}}}}

\DeclareRobustCommand{\unitsx}[1]{%
\ensuremath{\foreach \x/\y in {#1} {{\,\mathrm{\x}^{\y}}}}}

\DeclareRobustCommand{\chem}[1]{%
\ensuremath{\foreach \x/\y in {#1} {{\mathrm{\x}_{\y}}}}}

\newcommand{\simx}{\,{\sim}\,}

\newcommand{\kmo}{\chem{K/0.3,Mo/,O/3}}
\newcommand{\tas}{\ensuremath{
1T\text{-}\chem{Ta/,S/2}
}}
\newcommand{\Tc}{\usr{T}{c}{}}
\newcommand{\Tcz}{\usr{T}{c0}{}}
\newcommand{\tisa}{\ensuremath{{
\chem{Ti:/,Al/2,O/3}}}}
\newcommand{\feff}{\usr{f}{eff}{}}

\newcommand{\epsz}{\usr{\epsilon}{0}{}}
\newcommand{\epsr}{\usr{\epsilon}{r}{}}

\newcommand{\dsig}{\ensuremath{\Delta\sigma}}
\newcommand{\lamex}{\usr{\lambda}{ex}{}}
\newcommand{\Fex}{\usr{F}{ex}{}}
\newcommand{\Dex}{\usr{D}{ex}{}}
\newcommand{\absex}{\usr{\alpha}{ex}{}} 
\newcommand{\Nex}{\usr{N}{ex}{}}
\newcommand{\drrel}{\ensuremath{\Delta r/r_0}}
\newcommand{\drrelx}{\ensuremath{\Delta r(\nu,\tau)/r_0(\nu)}}

\newcommand{\hnux}{
\usr{h\nu}{ex}{}
}
\newcommand{\fsec}{\unit{fs}}
\newcommand{\psec}{\unit{ps}}
\newcommand{\thz}{\unit{THz}}
\newcommand{\ghz}{\unit{GHz}}
\newcommand{\eV}{\unit{eV}}
\newcommand{\nm}{\unit{nm}}
\newcommand{\mum}{\unit{\mu m}}

\newcommand{\pcmq}{\units{cm/-3}}
\newcommand{\muJcm}{\units{\mu J/,cm/-2}}
\newcommand{\kf}{\usr{k}{F}{}}
\newcommand{\Kel}{\unit{K}}

\newcommand{\DE}{\ensuremath{\Delta E}}
\newcommand{\nuLm}{\ensuremath{
\nu_{0n}
}}
\newcommand{\dnuLm}{\ensuremath{
\Delta\nu_{n}
}}
\newcommand{\nuLA}{\ensuremath{
\usr{\nu}{0A}{}
}}

\newcommand{\dnuLA}{\ensuremath{
\Delta\usr{\nu}{A}{}
}}

\newcommand{\Ommzsq}{\ensuremath{
\omega^{2}_{0n}
}}

\newcommand{\nuLmz}{\ensuremath{
\nu_{0n}^{(0)}
}}

\newcommand{\nuLmzz}{\ensuremath{
\nu_{0,n-1}^{(0)}
}}

\newcommand{\nuLAz}{
\usr{\nu}{0A}{(0)}
}

\newcommand{\epsbr}{\usr{\epsilon}{br}{}} 
 
\newcommand{\depsr}{\usr{\Delta\epsilon}{r}{}} 

\newcommand{\nup}{
\usr{\nu}{pl}{}
}

\newcommand{\meff}{
\usr{m}{eff}{}
}

\newcommand{\Delc}{\ensuremath{
\tilde\Delta
}}
\newcommand{\xic}{\ensuremath{
\tilde\xi
}}

\newcommand{\Uimp}{\usr{U}{i}{}}

\newcommand{\Wimp}[1]{
\usr{\Omega}{i}{#1}
}

\newcommand{\gimp}[1]{
\usr{\gamma}{i}{#1}
}

\newcommand{\ggimp}[1]{
\usr{g}{i}{#1}
}

\newcommand{\Rex}[1]{
\mathrm{Re}\{#1\}
}

\newcommand{\Imx}[1]{
\mathrm{Im}\{#1\}
}

\newcommand{\Eit}{\ensuremath{
\usr{E}{i}{}
}}
\newcommand{\Ert}{\ensuremath{
\usr{E}{r}{}
}}
\newcommand{\Ett}{\ensuremath{
\usr{E}{t}{}
}}
\newcommand{\Pit}{\ensuremath{
P\{\usr{E}{i}{}\}
}}
\newcommand{\Prt}{\ensuremath{
P\{\usr{E}{r}{}\}
}}
\newcommand{\Ptt}{\ensuremath{
P\{\usr{E}{t}{}\}
}}

\newcommand{\Pnt}{\ensuremath{
P\{E_n\}
}}

\begin{document}

\title{\papertitle}

\author{M.~D.~Thomson}
\author{K.~Rabia}
\author{F.~Meng}
\affiliation{\ffm} 
\author{M.~Bykov}
\author{S.~van Smaalen}
\affiliation{ Laboratory of Crystallography, University of Bayreuth, 95440 Bayreuth, Germany}
\author{H.~G.~Roskos\footnote{Corresponding author: Roskos@Physik.uni-frankfurt.de}} 
\affiliation{\ffm} 

\date{\today}
\begin{abstract} % Max 600 characters (including spaces)
We investigate the low-energy phase excitations of the quasi-one-dimensional charge density wave (CDW)
in \kmo, by direct probing of infrared-active CDW-lattice modes (phase-phonons) with ultrafast terahertz spectroscopy.
Both the nonequilibrium response and temperature dependence of the bands are reconciled by generalizing the time-dependent Ginzburg-Landau theory, beyond that previously applied to the amplitude-phonons, to include impurity effects, while the photoinduced blue-shifts are attributed to a reduction of the electron-phonon coupling induced by a long-lived free-carrier population.
\end{abstract}


\maketitle
The low-temperature CDW phase in organic and inorganic solids \cite{grun88} serves as an important prototype for a range of intriguing phenomena in systems with strongly coupled degrees of freedom and spontaneously symmetry-broken ground states  \cite{khom10}. 
The study of CDWs has taken on renewed relevance as they are found to arise as promoting/competing excitations in high-temperature superconductors \cite{chang12}.
The periodic modulation of both electron and lattice density (with $q=2\kf$) due to electron-phonon (e-ph) coupling is driven by a Peierls instability in 1D (or more generally, Fermi nesting in higher dimensions).
CDW systems exhibit low-energy excitations
which correspond to distortions (with wavevector $Q$) of the amplitude and phase of the CDW ground-state (GS) and whose quanta are a mixture of the CDW charges and bare phonons.
These modes ideally decouple into a (Raman-active) amplitude channel and (infrared-active, Goldstone) phase channel, which for $Q=0$ corresponds to a uniform amplitude scaling and spatial translation of the CDW, respectively.
While many studies focused on a single coupled phonon \cite{lee74,grun88}, which leads to an ``amplitudon'' and ``phason'' (the latter at $\nu{\sim}0$), in general multiple phonons can couple to the CDW electrons
yielding a set of amplitude- and phase-``phonons'' \cite{rice76,rice78,Degi91a,scha10,scha14}, 
which can be exploited as spectroscopic probes of the CDW physics.
The phase-modes appear in the optical conductivity even for centrosymmetric phonons due to the coupled CDW motion \cite{rice76}.

\begin{figure}
\includegraphics[width=0.9\columnwidth]{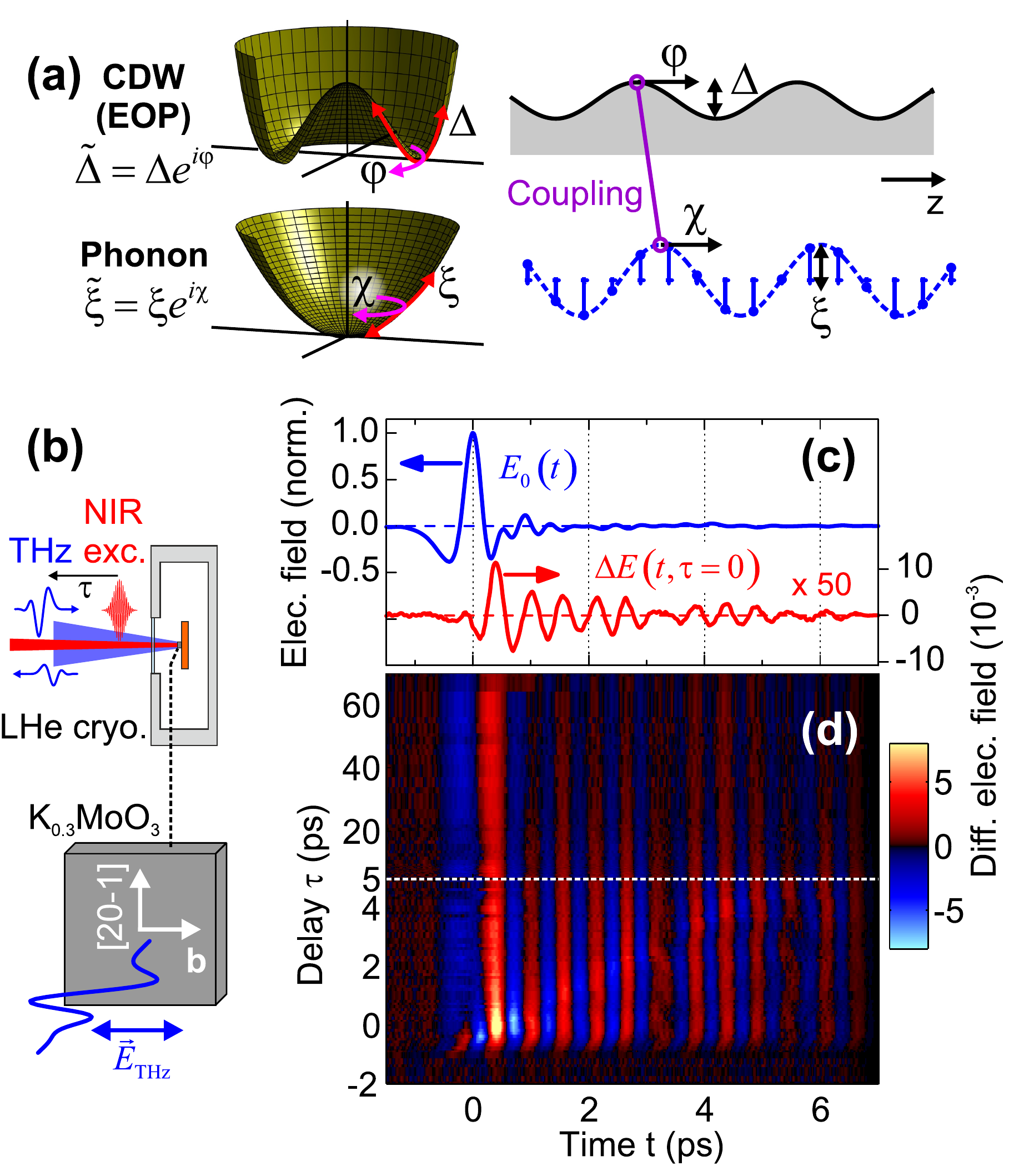}%
\caption{(a) Schematic of phase-space potentials (left) and spatial modulation (right) associated with electronic CDW (order parameter, EOP) and a single phonon, with wavevectors $q=2\kf$ and complex amplitudes $\Delc$ and $\xic$, respectively. 
(b) Experimental OP-TP reflection geometry. (c) Detected temporal electric fields ($T=50 \unit{K}$, pump fluence $\Fex = 550\muJcm$): 
Reference field $E_0(t)$ (reflected signal without optical excitation); differential field change $\DE(t,\tau)$ for pump-probe delay $\tau=0$. 
(d) Full set of measurements $\DE(t,\tau)$. Note that the data is plotted with two different scales for the ranges $\tau\in[-2,5]\psec$ and $\tau\in[5,70]\psec$, respectively.
\label{fig:timedomain}}
\end{figure}

Blue bronze ($\kmo$) is a quasi-1D conductor \cite{grunbook} which forms an incommensurate CDW below $\Tc=183\Kel$, and whose low-energy spectrum has been intensely studied via neutron scattering \cite{Sato83,Pou91}, Raman \cite{Trav83,Sag08}, and far-infrared \cite{Trav84,Degi91a} spectroscopy. 
More recently, the non-equilibrium response following femtosecond optical excitation has also been studied using optical \cite{dem99,sag07,tome09,scha14,yus10}, photoelectron \cite{liu13} and electron-diffraction \cite{hub14} probes.
In optical-pump optical-probe (OP-OP) experiments, the amplitude-phonons are detected via coherent oscillations in the interband permittivity.
Such data 
were employed to provide temperature-dependent band parameters for the amplitude-phonons \cite{scha10,scha14}, and were shown to be consistent with a time-dependent Ginzburg-Landau (TDGL) treatment of the GS with a linear coupling between the electronic order parameter (EOP, \Delc) and bare phonon ($\xic_n$) coordinates (Fig.~\ref{fig:timedomain}(a)).
This model yields a set of amplitude- and phase-modes similar to quantum mechanical (QM) models \cite{lee74,rice76,rice78}, and could account both for the $T$-dependence of several modes and the incomplete softening of the ``amplitudon'' band at $1.68\thz$ for $T\to \Tc$.
For the phase modes, while previous GS (far-)IR measurements covered a wide frequency range \cite{Trav84,Degi91a}, these were confined to low temperature, and hence data is lacking to further test the predictions of these models. 
Moreover, to our knowledge, no studies of the non-equilibrium response of the phase-phonons have been reported (although a signature assigned to the phason below $50\ghz$ was resolved in OP-OP measurements \cite{Xu06}).  Here we apply optical-pump THz-probe spectroscopy (OP-TP), which probes the phase-phonons via the broadband, transient \textit{complex} conductivity, and reconcile both the non-equilibrium dynamics and $T$-dependence of the bands with a generalized TDGL model.

The OP-TP experiments are described in detail in \cite{supplement}.   
Briefly, both optical-pump ($\hnux=1.6\eV$) and THz-probe pulses (time delay $\tau$) were normally incident on the sample (Fig.~\ref{fig:timedomain}(b)), and the reflected THz field was measured with time-domain sampling (frequency range $\simx 0.2\mbox{-}3\thz$).
While measurements were performed with the THz field polarized both parallel and perpendicular to the  $b$-axis, the phase-phonon signals presented here were observed only in the former case.
Also, measurements were made with fluence in the range $\Fex=140$-$720\muJcm$, although we present here only data for $\Fex=550\muJcm$ which yielded a favorable compromise between signal strength 
and saturation effects \cite{tome09}. 

\begin{figure}
\includegraphics[width=0.8\columnwidth]{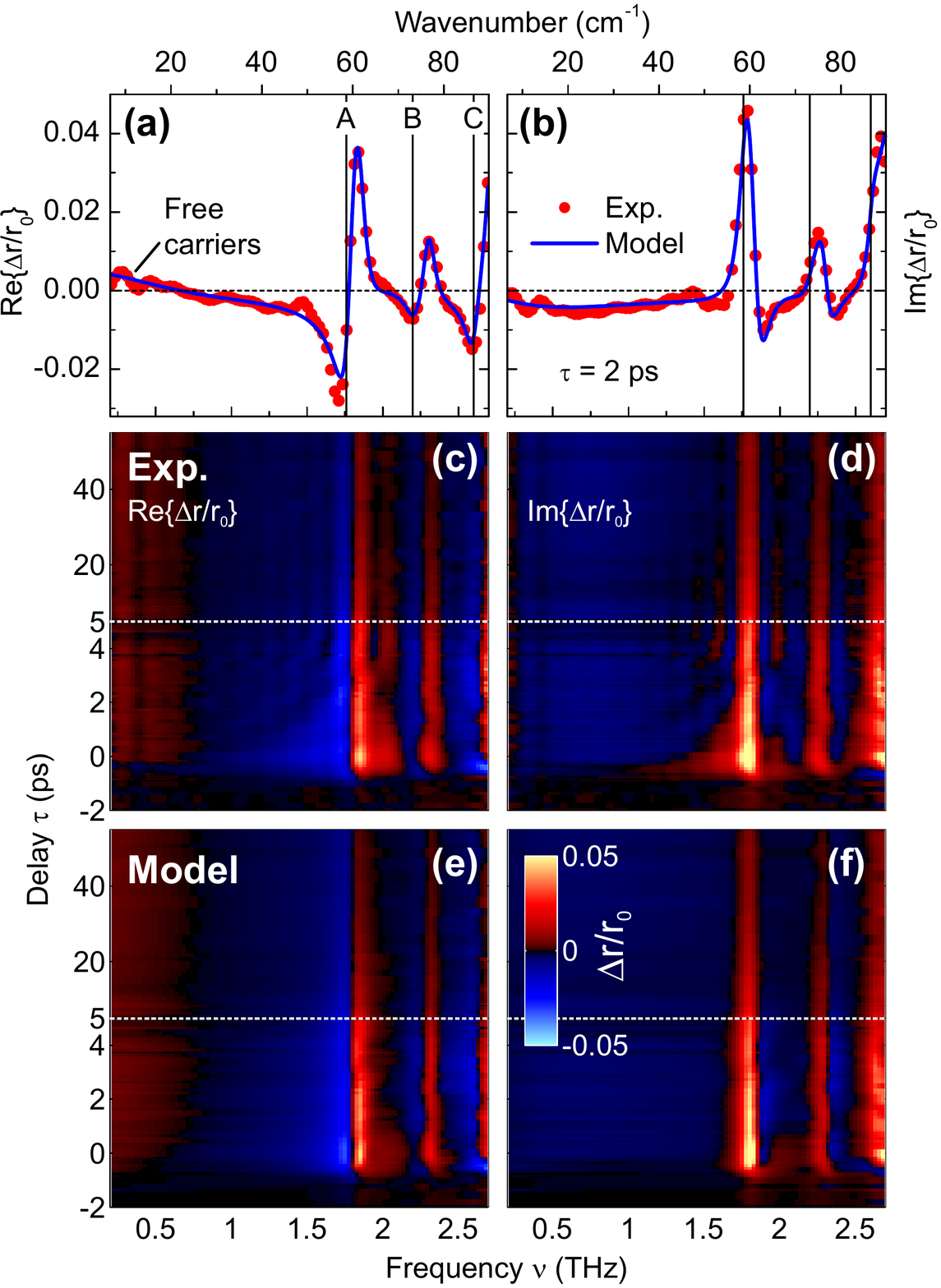}%
\caption{
Differential field reflectivity spectra $\Delta r/r_0\propto \dsig/(\epsr-1)$ at $T=50 \unit{K}$ (left graphs $\Rex{\drrel}$; right graphs $\Imx{\drrel}$).
(a,b) For pump-probe delay $\tau=2\psec$, including both experimental data (points) and fitted model (see text).  
The GS frequencies of the three main bands (labeled A-C) from the OP-TP analysis are included as vertical lines.
(c-d) Full experimental data  $\Delta r(\nu,\tau)/r_0(\nu)$ obtained from Fourier analysis of each row in Fig.~\ref{fig:timedomain}(d); (e,f) corresponding model fits.
\label{fig:drspec}}
\end{figure}

Examples of the reference reflected pulse $E_0(t)$ and differential signal $\DE(t,\tau=0)=E(t,\tau)-E_0(t)$ are shown in Fig.~\ref{fig:timedomain}(c) for $T=50\Kel$, while the full 2D signal is shown in Fig.~\ref{fig:timedomain}(d).
One sees that the temporal onset of $\DE$ is shifted somewhat from the main peak of $E_0$ and exhibits a ringing signature composed of multiple beating frequencies.
The corresponding differential reflectivity spectra $\Delta r(\nu,\tau)/r_0(\nu)$ are presented in Fig.~\ref{fig:drspec}(a-d).
Accounting for the short excitation depth $\Dex=1/\usr{\alpha}{ex}{}$ compared to the THz probe wavelengths, one can show \cite{supplement} that the complex conductivity change $\Delta\sigma(\nu,\tau)$ is essentially in-phase with $\Delta r/r_0$, and hence spectral features in $\drrel$ correlate closely with those in $\dsig$ (with minor distortions due to a pre-factor containing the GS permittivity).
The spectra in Fig.~\ref{fig:drspec}(a) contain a sequence of derivative-shape features in the range ${\gtrsim}1.5\thz$ for $\Rex{\drrel}$ ($\tau=2\psec$), whose polarity implies a photoinduced \textit{blue}-shift of existing bands. The most dominant feature (with a zero-crossing at $1.78\thz$) occurs about a frequency where 
a weak spectral modulation was indeed observed previously \cite{Degi91a} but not assigned as a phase-phonon.
From additional GS THz measurements and the analysis below, we deduce that the signatures arise from  
three main bands (labeled A-C from hereon, as shown in Fig.~\ref{fig:drspec}(a,b)), where bands (B,C) correspond to those previously found in FTIR measurements \cite{Degi91a}.
The weak signature at low frequency can be fit well with a Drude model (see below), which we assign to the generation of mobile carriers.  From the full 2D data (Fig.~\ref{fig:drspec}(c-d)) one sees that all features persist for delays beyond $50\psec$, after initial relaxation in the first few \psec.

\begin{figure}
\includegraphics[width=\columnwidth]{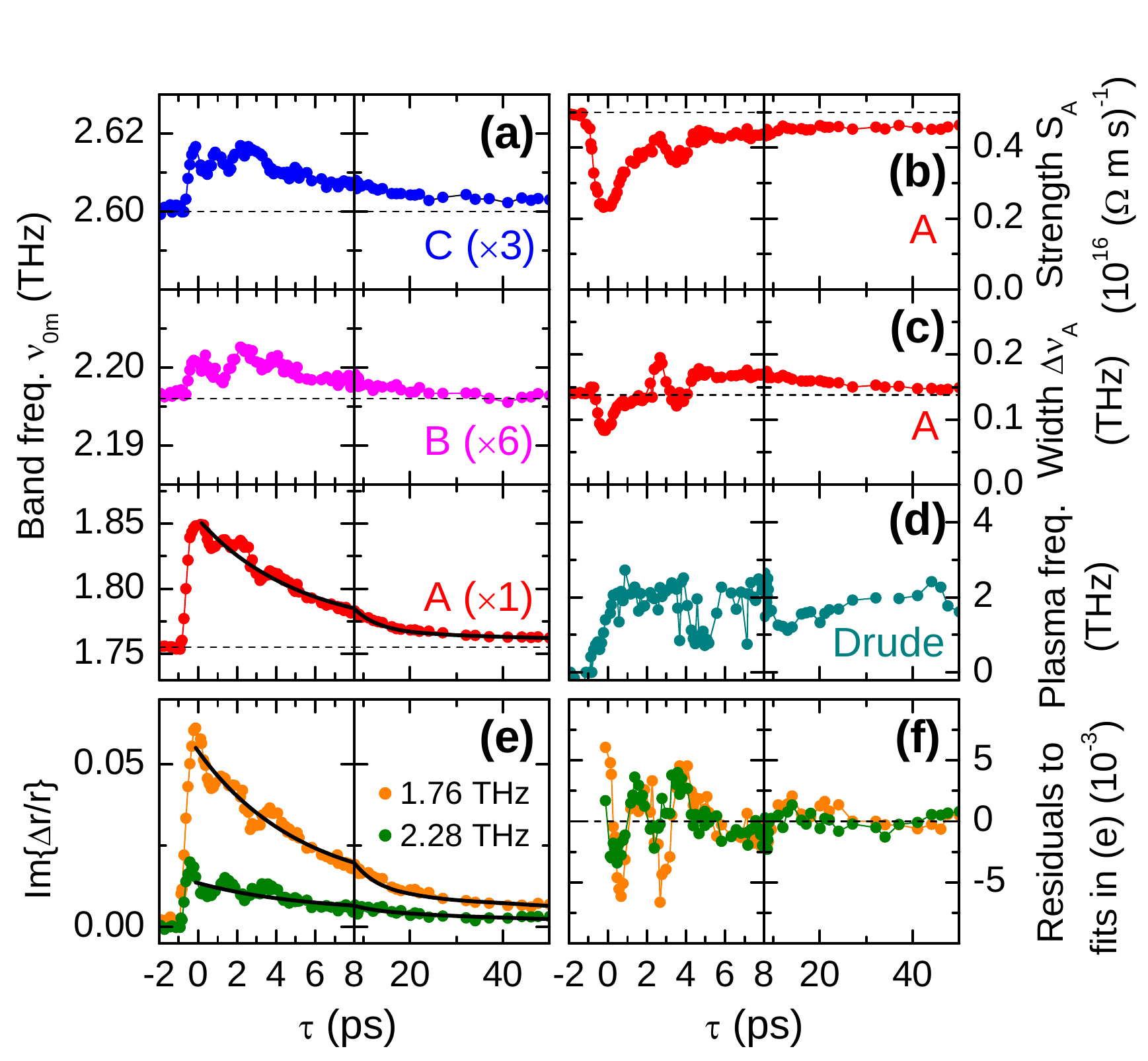}%
\caption{
Results of fitting $\drrel$ data in Fig.~\ref{fig:drspec} ($T=50\unit{K}$) vs. pump-probe delay $\tau$. 
(a) Band frequencies $\nuLm$ for the three bands $n=$A,B,C.  GS values obtained during global fit indicated by dashed horizontal lines.
(b,c) Band strength and bandwidth (FWHM) for Band A, respectively. (d) Fitted Drude plasma frequency $\nup$ for free-carrier contribution. 
Note that the fluctuations at later delays arise due to small drifts and noise during the measurement run, and any structure should rather be attributed to the non-monotonic order of delays used for the measurement run.
(e) Experimental kinetics of $\Imx{\drrel}$ at two selected frequencies, as indicated, and global bi-exponential fit (solid curves). (f) Corresponding residual from fitting in (e).
\label{fig:fits}}
\end{figure}

In order to disentangle the band contributions and extract parameter kinetics vs. $\tau$, we fit the full complex data based on photoinduced modification of three Lorentzian bands and a Drude contribution \cite{supplement}.
The resulting fitted spectra are shown in Fig.~\ref{fig:drspec}(e,f) (and also as curves in Fig.~\ref{fig:drspec}(a,b) for $\tau=2\psec$), which reproduce the experimental data well, except for the additional broadening seen in the first $2\psec$.  
Simulations for a single perturbed phonon \cite{supplement} show that this transient broadening can be attributed to the response of a time-non-stationary medium \cite{nien05} due to sub-ps dynamics, and hence cannot be reproduced using the quasi-stationary spectral analysis here.
In Fig.~\ref{fig:fits}(a-d) we plot selected fitted band parameters as a function of delay $\tau$, where the blue-shifts for all bands A-C are clearly evident ($\nuLm(\tau)$, Fig.~\ref{fig:fits}(a)).  
For band A, this is accompanied with a transient reduction in band strength ($\usr{S}{A}{}$, Fig.~\ref{fig:fits}(b)), and additional broadening (after initial development in the first few ps, Fig.~\ref{fig:fits}(c)).
As expected from inspection of the low-frequency region in Fig.~\ref{fig:drspec}(c), the fitted Drude plasma frequency rises to a fairly constant plateau at $\nup=2\thz$ (Fig.~\ref{fig:fits}(d)) with a   scattering time $\usr{\tau}{s}{}\simx 90\fsec$.  
Based on the low-frequency value of $\epsr(\nu\to 0)\simx 100$, $\nup$ corresponds to an excitation density $\Nex/(\meff/m_0)\simx5\times 10^{18}\pcmq$.  
As these photoexcited carriers may well populate orbitals distinct from those of the CDW condensate, we do not necessarily associate the effective mass here with that of the CDW 
($\usr{m}{eff}{CDW}\simx 300\cdot m_0$ \cite{grunbook}).

We note that the excitation depth $\Dex$ was also included in the fitting procedure (using a common value for all temperatures in the range $T=50\mbox{-}160\Kel$), and a value of $\Dex=3.8\mum$ was required to reproduce the signal strength and qualitative features, which is more than an order of magnitude larger than that expected from linear measurements ($\Dex<100\unit{nm}$ at $\lamex$ \cite{sag07}). 
One explanation for this could be strong bleaching of the excitation transition, especially given the high nominal excitation density $2.9\cdot 10^{20}\pcmq$ at the front surface.
However, from pump-only THz-emission experiments (to be dealt with in a subsequent report \cite{rabia16}), we see clear evidence for a photo-Dember-type effect, i.e. rapid diffusion of charges into the bulk due to the excitation gradient, which could lead to the larger effective excitation depth.

To extract kinetic time scales, we fit the experimental data directly, 
i.e. the kinetics of $\Imx{\drrel}$ for $\nu=1.76$ and $2.28\thz$ (the two peaks in Fig.~\ref{fig:drspec}(b)) as shown in Fig.~\ref{fig:fits}(e). A global bi-exponential fit yielded time constants of $\tau_1=5.1\psec$ and $\tau_2=60\psec$, in addition to a small but significant, long-lived offset.  The values $\tau_{1,2}$ also reproduce the band shift kinetics well (solid curve in Fig.~\ref{fig:fits}(a) for band A).
The fit residuals (Fig.~\ref{fig:fits}(f)) exhibit roughly two oscillation cycles with period $\simx 2\psec$ followed by a weaker oscillation with period $\sim 10\psec$.  This additional slow modulation of the system, which is in-phase for both bands, 
presumably results from coherent charge/lattice oscillations polarized perpendicular to the surface.
As discussed below, we attribute the blue-shifts to a reduction of the e-ph coupling due to the presence of the free carriers (i.e. a reduction of the red-shift of the dressed phonons).
As the plasma frequency $\nup(\tau)$ shows no such decay kinetics, this would imply that after the initial reaction of the phonons to the free charges, the system can reconfigure in such a way that the e-ph interactions can be partially reestablished. 
This process may again involve relaxation of distributions perpendicular to the surface. 
Compared to OP-OP studies \cite{sag07,scha10} of the amplitude-phonons 
(and interband electronic contributions), our value of $\tau_1$ is consistent with a component 
in the range $5$-$10\psec$ \cite{tome09,scha10,scha14} which was assigned to a second stage of CDW recovery, whereas the initial sub-ps CDW recovery (on the same time scale as the period of the THz waves here) would rather contribute to the transient broadening, as seen in our data.
A persistent offset in the (electronic) OP-OP signals was attributed tentatively to residual cooling \cite{sag07}, although our results here indicate that a long-lived free-carrier contribution
may well contribute.

\begin{figure}
\includegraphics[width=\columnwidth]{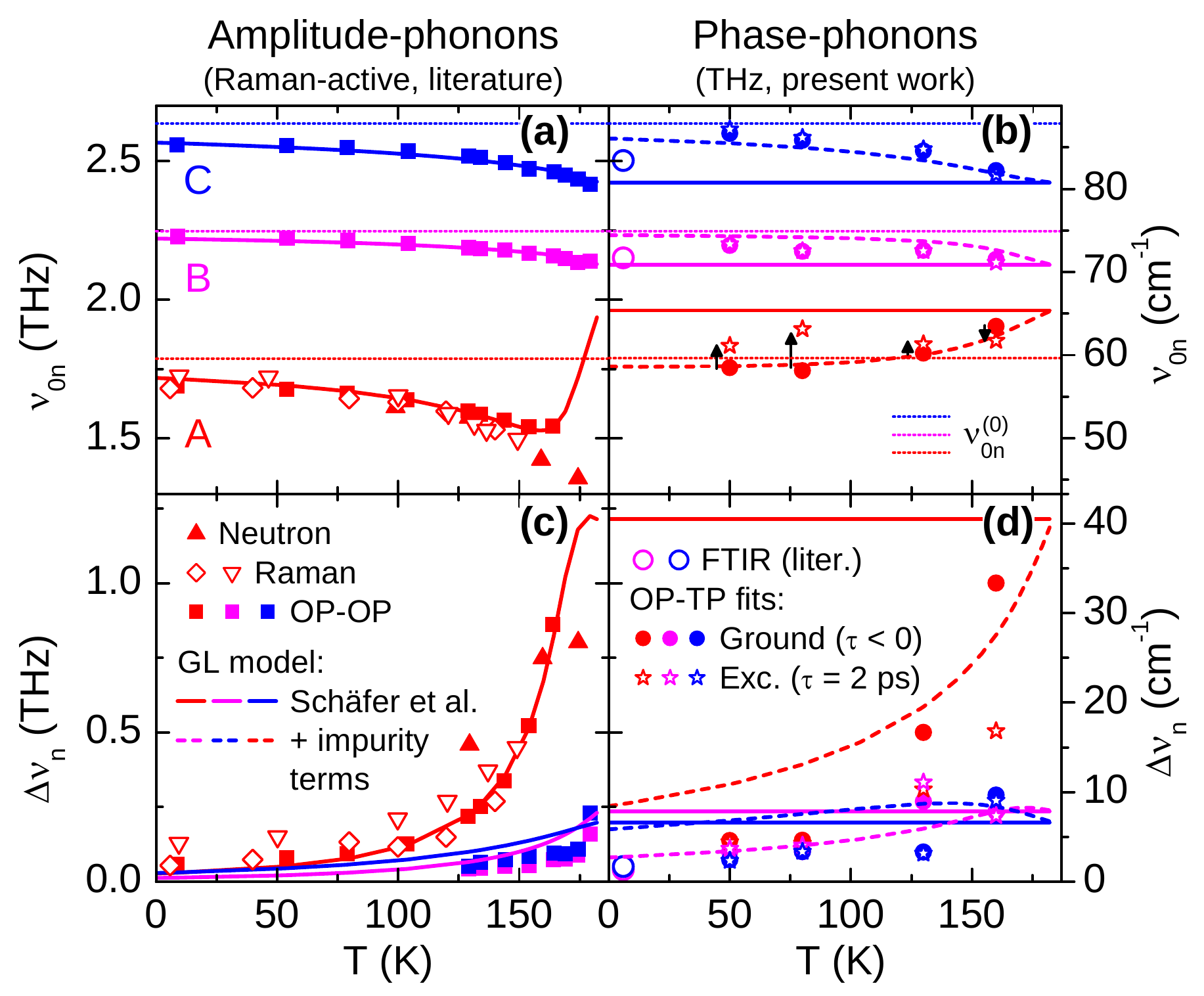}%
\caption{
Temperature dependence of amplitude- (left graphs) and phase-phonons (right) for $T<\usr{T}{c}{}$; center frequencies $\nuLm$ (top graphs) and bandwidths $\dnuLm$ (FWHM, bottom).
Literature data from 
neutron scattering ($\blacktriangle$ \cite{Pou91}),
Raman scattering ($\medtriangledown$ \cite{Trav83};
$\meddiamond$ \cite{Sag08}), 
optical-pump optical-probe reflectivity ($\blacksquare$ \cite{scha10,scha14}), 
and far-infrared FTIR measurements ($\medcircle$ \cite{Degi91a}, $T=6\unit{K}$).
THz TDS data (present work) from global analysis of OPTP data (for $T=50,80,130,160\unit{K}$) for both GS ({\large$\bullet$}) 
and $\tau=2\psec$ after excitation ($\medstar$).
Also included are model curves from the TDGL model (with parameters fit to the amplitude phonons in \cite{scha10,scha14} both without (solid) and with (dashed) pinning/scattering contributions from impurities (with $\Wimp{}(0)/2\pi=9.3\thz$ and $\ggimp{}=1.9$, see text). \label{fig:temperature}}
\end{figure}

We turn now to the temperature dependence of the phase-phonons, obtained from measurements as per those above for $T=50$, 80, 130 and 160$\Kel$ (see \cite{supplement}), which yielded the fitted band frequencies $\nuLm$ and widths $\dnuLm$ shown in Fig.~\ref{fig:temperature} (right) for both the GS, and excited-state  for $\tau=2\psec$. 
One sees a clear softening for band B and C, but rather a stiffening for band~A, with increasing $T$, accompanied with significant broadening for band~A.  The GS data for bands B, C from \cite{Degi91a} at $T=6\Kel$ are also included, and are reasonably consistent with an extrapolation of our parameters (recall that band~A was not assigned in \cite{Degi91a}). 

Literature data for the amplitude modes are also shown in Fig.~\ref{fig:temperature} (left), comprising GS neutron- and Raman-spectroscopy for band A, and OP-OP data for all three bands (sources cited in caption), which are reasonably consistent (except for the deviation between neutron and OP-OP data for $\usr{\nu}{0A}{}$ approaching $\Tc$).
Also included are the model curves based on the TDGL model adopted from \cite{scha10,scha14,supplement} which shows the very good agreement demonstrated by those authors (even for $T\ll\Tc$).
However, the nominal TDGL model predicts $T$-\textit{independent} phase-phonon bands (which coincide with the amplitude modes for $T\to\Tc$ -- solid horizontal lines in Fig.~\ref{fig:temperature}(b,d)), which  clearly is not consistent with our results away from $\Tc$.  While any GL treatment is only a first-order expansion about $T\approx\Tc$ \cite{khom10}, it is not clear why it should break down here specifically for the phase channel.  
Hence we considered possible $T$-dependent effects which should only affect the phase channel ($\Delta_2$, $\xi_{n2}$) in the TDGL equations.
As discussed in \cite{supplement}, a straightforward generalization which reproduces the observed trends for all bands is based on impurity effects, which dominantly affect the CDW \textit{phase} channel via pinning \cite{fuku78,tur94,wonn99} and scattering \cite{bak77,bak78}. We incorporated these via a pinning potential
$\Uimp=-\tfrac{1}{2}\Wimp{2}(T)\Delta_2^2$
and damping term $\gamma_2\to\gamma+\gimp{}(T)$, 
taken to depend on the relative equilibrium EOP amplitude
$\delta_0(T)=(1-T/\Tc)^{1/2}$
as per $\Wimp{2}(T)=\Wimp{2}(0)\delta_0^n(T)$
and $\gimp{}(T)/\gamma=\ggimp{}\delta_0^n(T)$,
where the exponent $n=2$ was required to obtain reasonable agreement with the data.
As can be seen in Fig.~\ref{fig:temperature}(b,d), the generalized model reproduces the qualitative trends in the phase-phonon frequencies $\nuLm(T)$ (including the crossing of $\nuLA$ through $\nuLAz$
at $T\simx 120\Kel$) and significantly improves the predicted trend for $\dnuLA(T)$, while maintaining all other TDGL parameters used for the amplitude-phonons.
The exponent $n=2$ deviates from the value $n=1$ which might be expected, e.g. the potential for a single impurity is usually taken as $\Uimp\propto\Wimp{2}\propto\delta_0$ \cite{tur94,fuku78,wonn99}.  
However, QM treatments have predicted impurity pinning forces which grow faster than $\Delta_0(T)$ \cite{tutto85,tuck89} due to competition between the CDW and formation of Friedel oscillations about each impurity. Also, a divergence in phase damping with decreasing $T$ was deduced from non-linear transport experiments \cite{flem86} and was discussed in terms of screening \cite{sned84} and impurity-induced phase deformations of the CDW, which enhance phase scattering as compensating thermal carriers are frozen out (which dominates over damping due to the corresponding CDW-carrier scattering which would instead grow with $T$ \cite{taka85}).

Considering the photoinduced band changes, one can see that at least qualitatively, the observed blue-shifts are consistent with a reduction of the couplings $m_n$ between EOP and bare phonons (shifting the frequencies $\nuLm$ in the direction of $\nuLmz$), which we attribute to the presence of the long-lived  free carriers. 
The fact that $\nuLA$ shifts even above $\nuLAz$ for $T<160\Kel$ (Fig.~\ref{fig:temperature}(b)) may be due to an additional partial suppression of the impurity pinning (recall that the solid lines are the predicted frequencies in the absence of pinning).
This raises the question whether the amplitude-phonons in OP-OP measurements are also blue-shifted \cite{scha10,scha14,yus10}. A study vs. fluence $\Fex$ \cite{tome09}, however, revealed a small \textit{red}-shift at high fluence.
One mechanism which could account for this different behavior involves relaxing the strict phase coherence of the coupling between EOP and bare phonons after excitation,   
$$U_n \to -m_n\Delta\cdot\xi_n [ (1-\eta) \cos(\varphi-\chi) + \eta], \quad (0 \leq \eta \leq 1)$$
e.g. due to the presence of inhomogeneous distortions of the CDW.  As shown in \cite{supplement}, this modification only reduces the effective coupling for the phase channel, consistent with the experimental findings.

In conclusion, the use of coherent THz pulses to probe the complex conductivity and phase-phonons in \kmo{} provides unique insight into the underlying CDW physics, such as impurity and screening effects which predominantly affect the phase channel.  The inclusion of these effects within a generalized TDGL model can account for the observed $T$-dependence and photoinduced dynamics.
Given that the TDGL model provides a highly practical framework for interpretation of (non-equilibrium) experiments \cite{scha10,scha14,yus10}, we strongly advocate further investigations into its applicability  and correspondence to microscopic QM treatments.

We acknowledge Alfred Suttner for the growth of the the \kmo{} single crystals; SvS acknowledges support
from the DFG under contract No. Sm 55/25-1.

%\begin{acknowledgments}
%\end{acknowledgments}

%\bibliography{kmo_optp_bib1}


%%%%%%%%%%%%%%%%%%%%%%%%%%%%%%%%%%%%%%%%%%%%%%%%

%\newpage
\clearpage

%\include{kmo_optp_arxiv_suppl}

%\documentclass[aps,prl,reprint,superscriptaddress]{revtex4-1}
%\documentclass[aps,rse,superscriptaddress,preprint,raggedbottom,notitlepage,11pt,tightenlines]{revtex4-1}

%\input{kmo_optp_header}
%\usepackage{placeins}

%\begin{document}

%\section{Supplementary material}
%\widetext
\begin{center}
\textbf{\large SUPPLEMENTAL MATERIAL}
\end{center}

\setcounter{equation}{0}
\setcounter{figure}{0}
\setcounter{table}{0}
\setcounter{page}{1}
\makeatletter
\renewcommand{\theequation}{S\arabic{equation}}
\renewcommand{\thefigure}{S\arabic{figure}}
\renewcommand{\bibnumfmt}[1]{[S#1]}
\renewcommand{\citenumfont}[1]{S#1}

%\section*{Supplementary material to: ``\papertitle''}

\section{Experimental details}

The single crystals of \kmo were grown according to the temperature gradient flux method \cite{S_Rama84}, reaching facial dimensions up to $4\times 5\units{mm/2}$, and the crystal structure and orientation were confirmed by x-ray diffraction on a small single crystal. The crystal surfaces were sufficiently flat/smooth to be used in the experiments as-grown.

\begin{figure}[b]
\centering
\includegraphics[width=\columnwidth]{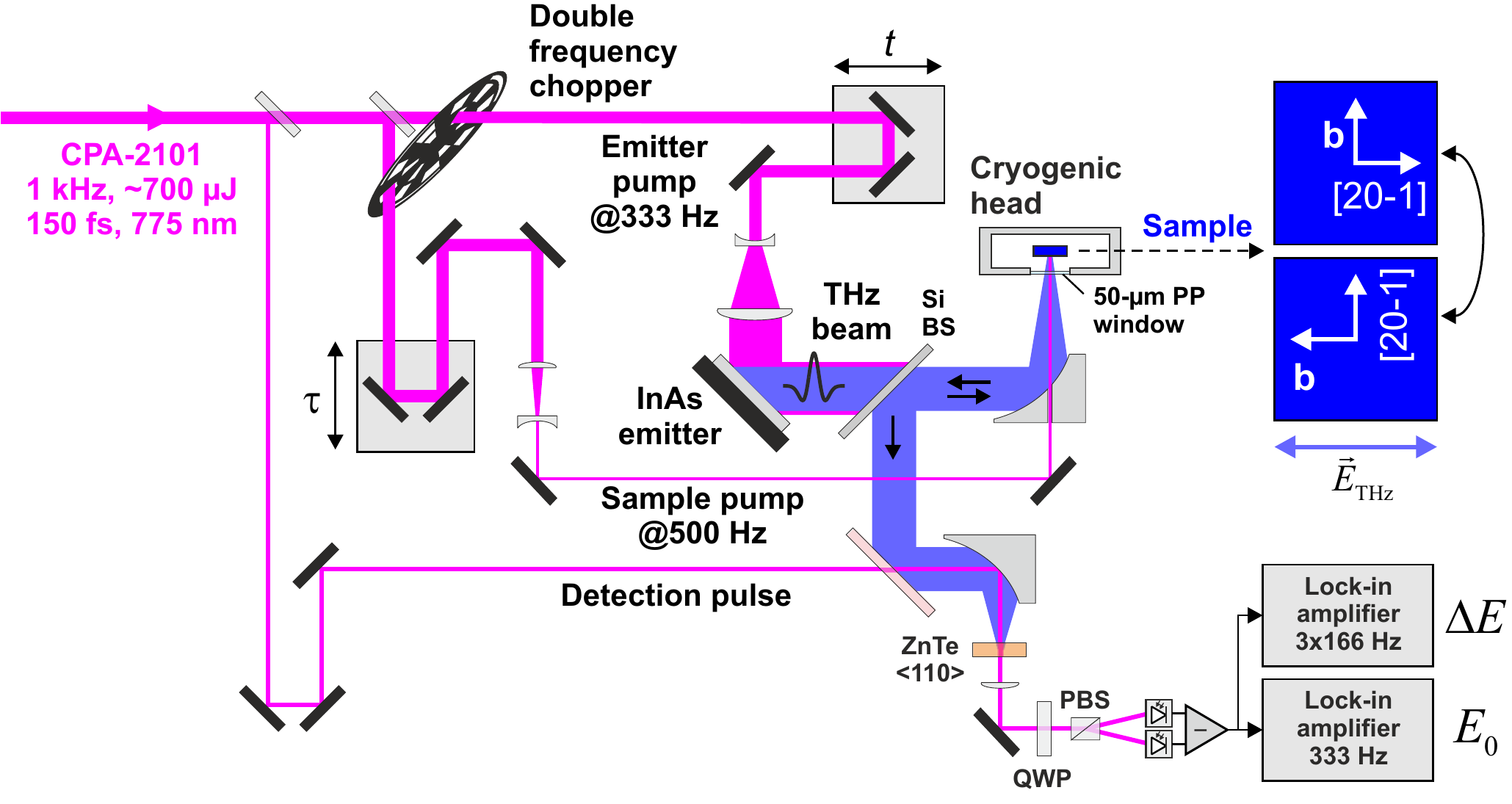}
\caption{Schematic of full experimental setup for optical-pump THz-probe measurements.
\label{fig:setup}}
\end{figure}

The terahertz time-domain spectroscopy system used for the optical-pump THz-probe experiments is depicted in Fig.~\ref{fig:setup}. The system is based on a 1-kHz \tisa{} amplifier laser (Clark-MXR CPA-2101, $\lamex=775\nm$, pulse duration $\simx 150\fsec$ FWHM).  The THz pulses were generated by a large-area InAs surface emitter, while a 0.5-mm-thick ${<}110{>}$-cut ZnTe is used for electrooptic detection.

The blue bronze samples were mounted in a liquid-helium cryostat (Oxford Microstat) and could be oriented with the conductive b-axis either parallel or perpendicular to the p-polarized THz probe field.
A thin polypropylene film (50-\mum-thick, Goodfellow PP301350) was used as the window to minimize any distortion of the THz beam (and two-photon absorption effects for the optical excitation pulse).
In order to realize a normal-incidence reflection geometry, a two-way beamsplitter (0.5-mm-thick high-resistivity Si wafer) is placed in the THz beam before the sample. The optical excitation beam for the sample passes through a small hole in the sample focusing mirror (off-axis paraboloidal mirror, OAPM, effective focal distance $\feff=101.6\unit{mm}$), brought to a weak focus on the sample (with a spot diameter of  $\simx 2\unit{mm}$) using a telescope.

The mechanical delay stage for THz detection was placed in the emitter arm, such that the time-variable $t$ corresponds to a fixed pump-probe delay $\tau$ for each point in the acquired THz waveform \cite{S_nemec02}. 
Besides yielding a 2D pump-probe signal $\DE(t,\tau)=E(t,\tau)-E_0(t)$ with a synchronous zero-delay for all $t$, this also rejects any pump-only emission signals in the resulting spectral data $\DE(\nu,\tau)$ (which were also observed, and will be addressed in a future report).
A dual-chopping scheme with two lock-in amplifiers was employed to synchronously measure $\DE$ and $E_0$ \cite{S_Iwa09}, which was essential to track phase drifts in the reflected THz field due to unavoidable longitudinal movement of the cryostat (of some $10\mum$).

\section{Derivation of approximate relation between differential reflectivity and conductivity for an exponential excitation profile}

The complex field reflection coefficient for a sample where the excitation depth is small compared to the probe wavelength (as is the case for blue bronze with near-infrared excitation and THz probing) deviates significantly from the Fresnel formula for a homogeneous medium \cite{S_Vine84,S_meng15}.
For a pump-induced change in the sample complex permittivity $\depsr(\omega,z)=\depsr(\omega)e^{-z/\Dex}$ (where $\Dex=1/\absex$ is the excitation depth) the reflectivity can still be expressed in the form
$r=(1-n')/(1+n')$ 
where $n'=n+\sqrt{\depsr}X_\beta(\xi)$, 
$n=\sqrt{\epsr}$ is the background (unpumped) refractive index, and
$X_\beta(\xi)=I_{\beta+1}(\xi)/I_{\beta}(\xi)$ is the ratio of modified Bessel functions with 
$\beta=2i\omega n \Dex/c$ 
and $\xi=2i\omega\sqrt{\depsr}\Dex/c$.
For $\Dex\ll \usr{\lambda}{THz}{}$, one can approximate the expression for $X_\beta(\xi)$ to first-order as:
$X_\beta(\xi)\to \tfrac{1}{2}\xi/(\beta+1)$.
Substituting this into the expression for $r$, noting that the ground-state reflectivity $r_0=(1-n)/(1+n)$, one can readily derive (assuming $\Delta n=n'-n\ll n$):
\begin{equation}
\frac{\Delta r}{r_0}
\approx \frac{2\Dex}{1+2ik_0 \Dex \sqrt{\epsr}}\frac{1}{\epsz c (\epsr-1)}\Delta\sigma,
\label{eq:drrel}
\end{equation}
where $\Delta r = r-r_0$, $k_0=\omega/c$, and $\Delta\sigma=i\epsz\omega\depsr$ is the pump-induced change in complex conductivity.  The validity of these approximations was thoroughly checked in numerical tests with parameters appropriate to the experiments presented in the main paper.
Note that if one were instead to use a simple Fresnel reflection treatment assuming homogeneous excitation vs. depth, this introduces an imaginary unit in the coefficient relating $\drrel$ and $\dsig$, such that real and imaginary parts would be erroneously exchanged in calculating $\dsig$ from the data.

%\newpage

\section{Details of fitting with global ground-state parameters and OP-TP spectra for all temperatures}
The fit model for each complex field reflectivity spectrum in the OP-TP data $\drrelx$ was based on assuming three Lorentzian bands and a Drude contribution (as well as a frequency-independent background permittivity $\epsbr$ to account for higher-lying resonances), whose ground-state (GS) parameters (frequency $\nuLm$, width $\dnuLm$ and strength $S_n$) are modified by the optical excitation to $\nuLm'(\tau)$, 
$\dnuLm'(\tau)$, $S_n'(\tau)$ and 
$\usr{\epsilon}{br}{\prime}(\tau)$.
By definition, $\usr{\nu}{0,D}{} \equiv \usr{\nu}{0,D}{\prime} \equiv 0$ for the Drude contribution, and we assumed no free carriers in the GS ($\usr{S}{D}{}=0$).
Note that we also tested the use of Fano lineshapes (as was applied in the vicinity of a single band in an OP-TP study on \tas{} \cite{S_dean11}), especially as this could be appropriate for the proposed interaction with free carriers.  However, the Fano interaction (which essentially mixes the real and imaginary parts of the band conductivity) could not reproduce such large shifts, especially for band A.

Due to baseline/resolution issues in the GS THz-TDS results (inset in Fig.~2(a) in main paper), we included global GS band parameters in the fit procedure (although the fitted GS band strengths were constrained to remain comparable to those from the GS TDS data).  In order to implement a robust and practical fitting scheme including the GS parameters, we first selected spectra from a small number of delays ($\tau_n=2, 10\psec$), and performed the fitting of all parameters.  Thereafter, the GS parameters were held fixed and each spectrum was fitted sequentially for the excited-state (ES) parameters.
At each iteration, we first used the expression above (Eq.~\ref{eq:drrel}) to estimate the model differential field reflectivity $\drrel$.  
While Eq.~\ref{eq:drrel} is nonlinear in the GS complex permittivity, it is linear in the differential change.
This allowed us to determine the ES band strengths $S_n'(\tau)$ via generalized regression, which relieved the fitting algorithm of a fraction of the search parameters and facilitated the fitting procedure.  Then the exact expression for $\drrel$ was calculated and used to compute the current misfit.
To avoid stagnation, the nonlinear optimization was based on Covariance Matrix Adaptation Evolution Strategy (CMA-ES).
Note that the excitation depth $\Dex$ was also fitted (as the nominal value from linear measurements could not account for the experimental data, see main paper).  However, after the optimum value was determined with the data for $T=50\Kel$, it was kept fixed for all other temperatures $T=80,130,160\Kel$.

\begin{figure}[h]
\centering
\includegraphics[width=\columnwidth]{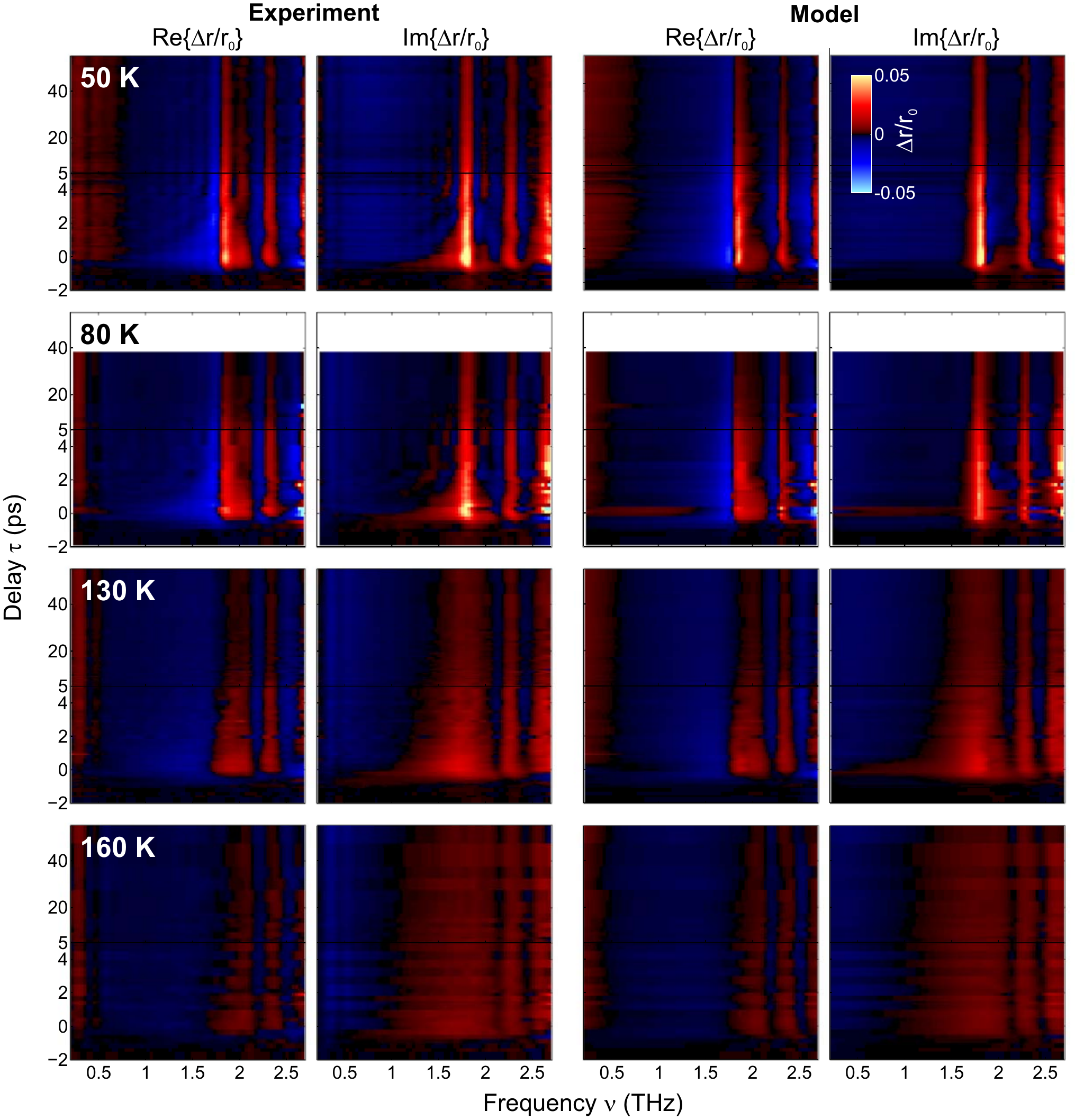}
\caption{Transient reflectivity spectra for all measured temperatures $T=50, 80, 130, 160\Kel$ with excitation fluence $\Fex = 550\muJcm$ (as per data in main paper for $T=50\Kel$ (Fig.~2)). Experimental data (left two columns) and Lorentzian-Drude fits (right two columns); real and imaginary parts of $\drrel$ as indicated. 
\label{fig:tdspec}}
\end{figure}

The experimental and fitted spectra for all measured temperatures are shown in Fig.~\ref{fig:tdspec}

%\newpage
%\section{Complete set of transient spectra and fits for all temperatures}

%\FloatBarrier 

\section{Application of the time-dependent Ginzburg-Landau treatment}

Here we consider generalizations of the time-dependent Ginzburg-Landau (TDGL) model prescribed in 
\cite{S_scha10,S_scha14}.  
For completeness, we first briefly summarize this, in order to clarify the notation and modifications that follow.
We consider the potential function, in terms of the complex electronic order parameter 
$\tilde\Delta = \Delta e^{i\varphi} = \Delta_1 + i\Delta_2$ 
and complex bare phonon coordinates
$\tilde\xi_n = \xi_n e^{i\chi_n} = \xi_{n1} + i\xi_{n2}$ ($n=1\dots N$) (where all coordinates refer to the complex envelope amplitudes of the $q=2\kf$ components) as per:
\begin{equation}
U(\tilde\Delta,\tilde\xi_1,\dots,\tilde\xi_N)=U_{\Delta} + U_{\xi n} + \usr{U}{c}{}
\label{eq:potfunc}%
\end{equation} %
where $U_{\Delta}=-\tfrac{1}{2}\alpha(\Tcz-T)\Delta^2 + \tfrac{1}{4}\beta \Delta^4$ is the Mexican hat potential, 
$U_{\xi n}=\tfrac{1}{2}\Ommzsq \xi_n^2$ represents the elastic energy stored in the bare phonon mode $n$ with frequency $\omega_{0n}$, and
$\usr{U}{c}{}=-m_n(\Delta_1\xi_{n1}+\Delta_2\xi_{n2}) = -m_n\Delta\cdot\xi_{n}\cos(\varphi-\chi_n)$ is the linear coupling term (summations over $n$ are left implicit).
This has the equilibrium solution $\Delta_0^2=\frac{\alpha(\Tc-T)}{\beta}$ and $\xi_{0n}=\frac{m_n}{\Ommzsq}\Delta_0$, where $\Tc=\Tcz+\frac{m_n^2}{\alpha\Ommzsq }$ is the renormalized critical temperature.

Taking $\varphi_0=\chi_{n0}=0$ as the equilibrium phase (at first, arbitrarily, as there is no pinning term and hence $U$ depends only on $(\varphi-\chi_n)$), and calculating the Hessian matrix of the potential about $\tilde\Delta_0=\Delta_0$ yields the linearized equations of motion \cite{S_scha10,S_scha14}:
\begin{subequations}
\begin{align}
\partial_t^2{\hat\Delta_1}&=-\left[ 2\alpha(\Tc-T) + \frac{m_n^2}{\Ommzsq} \right]\hat\Delta_1
 +m_n\hat\xi_{n1} - \gamma_1\partial_t{\hat\Delta_1}\\
\partial_t^2{\hat\Delta_2}&=-\frac{m_n^2}{\Ommzsq}\hat\Delta_2
 +m_n\hat\xi_{n2} - \gamma_2\partial_t{\hat\Delta_2}\\
\partial_t^2{\hat\xi_{n1}}&=m_n\hat\Delta_{1} - \Ommzsq \hat\xi_{n1} \\
\partial_t^2{\hat\xi_{n2}}&=m_n\hat\Delta_{2} - \Ommzsq \hat\xi_{n2}
\end{align} \label{eq:motion}%
\end{subequations}
where $\hat\Delta_1\approx \Delta-\Delta_0$ and $\hat\Delta_2\approx\Delta_0 \varphi$ represent the amplitude and phase deviations from equilibrium (likewise for $\hat \xi_{n1}$,$\hat \xi_{n2}$), and we have added phenomenological damping constants $\gamma_{1,2}$ for $\hat\Delta_{1,2}$. Note that while in a classical treatment one has $\gamma_1=\gamma_2=\gamma$ \cite{S_bhatt75}, we allow here for $\gamma_1 \neq \gamma_2$ to account for different quasi-particle scattering channels. Moreover, the variables are normalized to unit mass and we do not include any inherent damping for the bare phonon modes (which did not assist in fitting the phase-phonon data). The equations for the amplitude (Raman-active) and phase (IR-active) channels are decoupled, and converge to the same set of solutions for $T\to\Tc$ (corresponding to the $T$-independent phase channel).

In the overdamped limit for the EOP (as adopted in \cite{S_scha10,S_scha14}), i.e. $\partial_t^2\Delta_j \ll \gamma \partial_t\Delta_j$ (which is equivalent to neglecting the inertial mass), Eq.s~\ref{eq:motion}(a-b) become:
\begin{subequations}
\begin{align}
\partial_t{\hat\Delta_1}&=-\kappa_1 \left[ 2\alpha(\Tc-T) + \frac{m_n^2}{\Ommzsq} \right]\hat\Delta_1
 +\kappa_1 m_n\hat\xi_{n1} \\
\partial_t{\hat\Delta_2}&=-\kappa_2\frac{m_n^2}{\Ommzsq}\hat\Delta_2
 +\kappa_2 m_n\hat\xi_{n2}
\end{align} \label{eq:motionod}%
\end{subequations}
where $\kappa_j=\gamma_j^{-1}$ is the scattering time.
As per \cite{S_scha10,S_scha14}, one can solve these linear equations with the ansatz $\propto e^{\lambda t}$ for the eigenvalues $\lambda_k=-\Gamma_k/2+i\Omega_{0k}$ (and corresponding eigenvectors, which reflect the relative contribution of the EOP and each phonon to the mode), which yields either $N+2$ (general damping, Eq.~\ref{eq:motion}) or $N+1$ (overdamped limit, Eq.~\ref{eq:motionod}) modes for each channel (after discounting for the complex-conjugate solutions with $\Omega_{0k}<0$).
For each channel, one obtains a set of $N$ finite-frequency modes which, at least for the parameters used for \kmo{} in \cite{S_scha10,S_scha14}, each contain a dominant contribution from one of the bare modes, and hence have frequencies $\Omega_{0k}$ only moderately shifted from a given bare phonon at $\omega_{0n}$ (although in general, stronger mixing and avoided crossings can result \cite{S_scha14}).
In the overdamped limit, the remaining eigenvalues have $\Im(\lambda)=0$, and also $\Re(\lambda)$ for the phase channel (and amplitude channel for $T\to \Tc$).  The latter was associated in \cite{S_scha10,S_scha14} with a true soft mode of the system \cite{S_khom10}. 
As the authors asserted, this indeed represents a distinct physical picture to that from earlier quantum mechanical (QM) treatments, as discussed below.

In order to provide a starting point for the analysis of our phase-phonon data vs $T$, we first 
re-fitted the model curves for the Raman band data (Fig.~3(a,c) in main paper), i.e. the frequencies and bandwidths extracted from \cite{S_scha10} (note that only the fitted values of the parameters $\kappa m_n^2$ were given there).  This yielded a fitted set of parameters as shown in Table~\ref{tab:glpars}, where the values of $\kappa m_n^2$ are in agreement with those given in \cite{S_scha10}.
\begin{table}
\def \colwidA{0.1\columnwidth}
\def \colwidB{0.06\columnwidth}
\begin{tabular}{
L{\colwidA} 
L{\colwidA} 
R{\colwidB} 
R{\colwidB} 
R{\colwidB}
}
\hline
 &  & \multicolumn{3}{c}{$n$} \tabularnewline 
 &  & \multicolumn{1}{c}{1} & \multicolumn{1}{c}{2} & \multicolumn{1}{c}{3} \tabularnewline 
\hline
$\omega_{0n}/2\pi$ & ($\mathrm{THz}$) & 1.79 & 2.25 & 2.64 \tabularnewline 
$\kappa m_n^2$ & ($10^{36} \units{s/-3}$) & 580 & 320 & 1150 \tabularnewline 
\hline
$\gamma_1^{-1}=\gamma_2^{-1}$ & ($\mathrm{fs}$) & \multicolumn{3}{c}{3.4} \tabularnewline 
$\alpha$ & \multicolumn{1}{c}{($\unitsx{ps/-2,K/-1}$)} & \multicolumn{3}{c}{46} \tabularnewline 
\hline
\end{tabular}
\caption{Fitted parameters TDGL parameters used both to reproduce the Raman channel data from \cite{S_scha10} and as the basis for the IR-active data measured in the present paper.}
\label{tab:glpars}%
\end{table} 

\subsection{Phenomenological impurity pinning/scattering vs. temperature}

As discussed in the main paper, the nominal TDGL model summarized above predicts temperature-independent band parameters for the phase-phonons, whereas a clear $T$-dependence is observed in the experimental THz data (Fig.~4(b,d), main paper).
Hence several physical generalizations/modifications of the model were investigated, in order to reconcile this $T$-dependence, which would still retain the good agreement with the literature amplitude-phonon data. This included (i) general damping (c.f. Eq.~\ref{eq:motion}, i.e. including the underdamped regime); (ii) a temperature dependence for $\beta$ -- although this does not actually affect the $T$-dependence of the phonon data, rather only the magnitude of the equilibrium coordinates and renormalization of the critical temperature $\Tcz\to\Tc$. (iii) Our efforts to generalize the TDGL potential to include gradient terms of the complex order parameter (as suggested in \cite{S_scha10}), in particular the term $|\partial_z \tilde\Delta|^2$ \cite{S_scal72,S_mckenz95}, did not yield any coupling between the amplitude- and phase-channels (at least to first-order, once reducing the spatial equations to the $q=2\kf$-components $\tilde\Delta$ and $\tilde\xi_n$). 
(iv) A finite pinning potential for the CDW phase \cite{S_tuck89,S_mckenz95,S_wonn99}, which was omitted in \cite{S_scha10,S_scha14} presumably as it should not strongly affect the amplitude channel and hence was not necessary to treat the behavior of the Raman-active bands.
(v) $T$-dependent damping $\gamma_2$ for the EOP phase ($\Delta_2$).  In this case, one could consider two opposite trends: i.e. $\gamma_2(T)$ \textit{increasing} with $T$, e.g. due to friction (scattering) with an increasing number of normal carriers thermally excited across the CDW gap \cite{S_mcmill75}; $\gamma_2(T)$ \textit{decreasing} with $T$, e.g. due to the loss of thermal carriers which may screen impurity/Coulomb interactions between charges in the CDW condensate and suppress the associated scattering processes \cite{S_bak77,S_bak78}.

\begin{figure}[h]
\centering
\includegraphics[width=0.9\columnwidth]{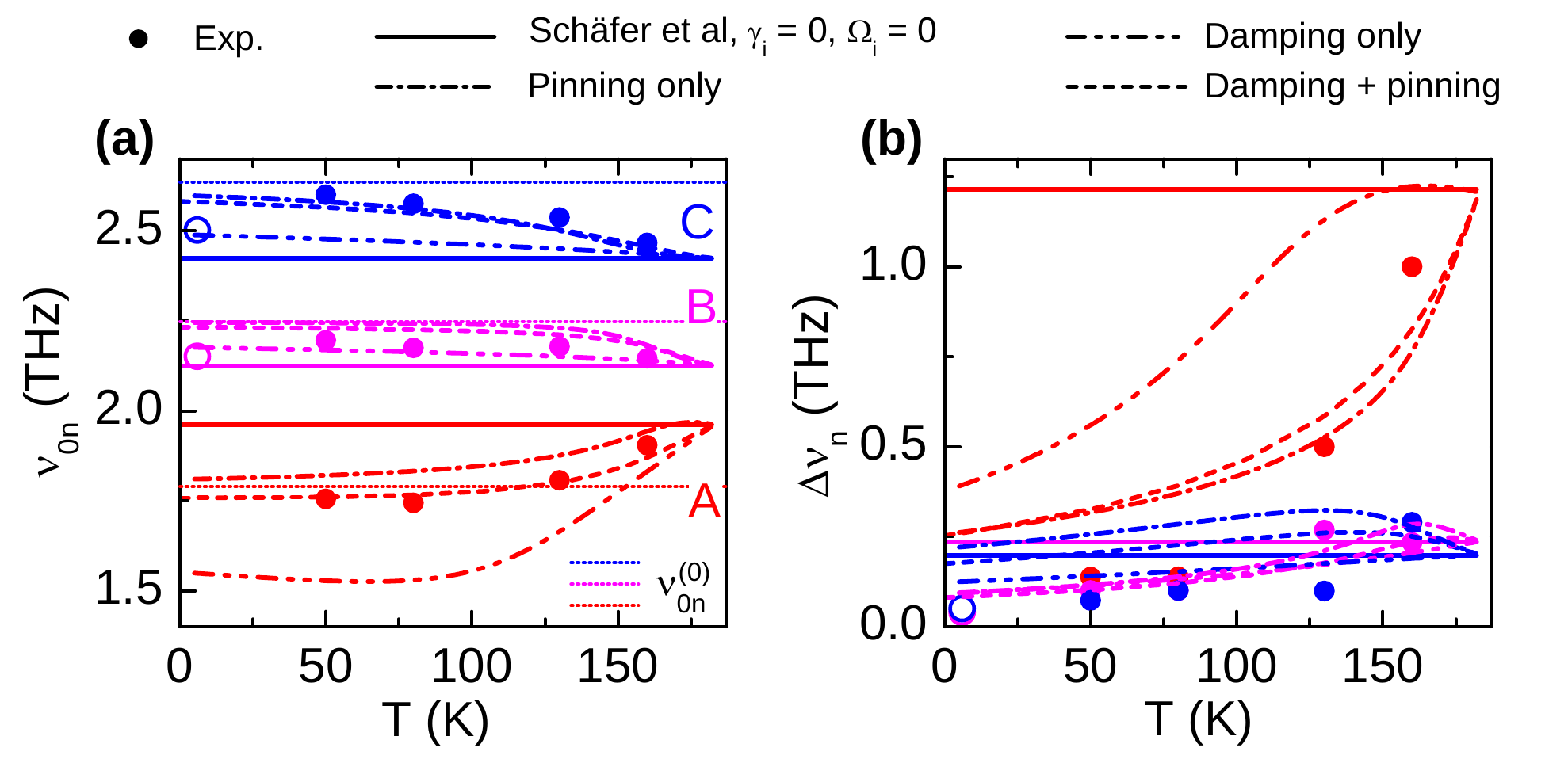}
\caption{Phase-phonons band parameters vs. $T$ from experimental and TDGL model with impurity damping and pinning: (a) frequencies $\nuLm$ and (b) bandwidths $\dnuLm$.  Description of the model parameters given in the text.  To illustrate the impact of each effect separately, we plot the model curves in all four cases: no effects (the nominal TDGL model from \cite{S_scha10,S_scha14}), pinning only, damping only, and both pinning/damping (as indicated in legend).
\label{fig:thzeffect}}
\end{figure}

We found that including effects (iv) and (v) (with $\gamma_2$ decreasing with $T$) allowed one to reproduce the (semi-quantitative) behavior of the phase-phonon band data vs. $T$, i.e. with band A stiffening (and crossing its bare mode frequency $\nuLAz$ at intermediate $T$) while bands B,C soften as $T\to\Tc$, and a significant reduction in the bandwidth of band A at low $T$.
These effects were implemented as follows.  For the impurity pinning, we treat a single strong-pinning center (which could be the consolidated average of several impurity potentials in each local region) and add to the function $U$ (Eq.~\ref{eq:potfunc}) a harmonic potential  
$\usr{U}{i}{}=\tfrac{1}{2}\Wimp{2}(T)\Delta_2^2$
\cite{S_tur94,S_wonn99} (taking the origin of the impurity at $\varphi=0$).
As both impurity pinning and scattering should depend on the (equilibrium) amplitude of the EOP, we took both the restoring force $\Wimp{2}(T)$ and damping $\gamma_2(T)$ to depend on $\Delta_0$, i.e. 
$\Wimp{2}(T)=\Wimp{2}(0)\delta_0^n(T)$ and $\gamma_2(T)=\gamma+\gimp{}(T)$ with $\gimp{}(T)/\gamma=\ggimp{}\delta_0^n(T)$, where 
$\delta_0(T) \equiv \Delta_0(T)/\Delta_0(0)=(1-T/\Tc)^{1/2}$ is the normalized EOP amplitude.
Tests showed that the value $n=2$ for the exponent was required to obtain reasonable agreement with the data (the physical significance of this is discussed in the main paper).
The best fit of the phase-phonon data corresponded to the values $\Wimp{}(0)/2\pi=9.3\thz$ and $\ggimp{}=1.9$ (using the same values for the other parameters above, and hence maintaining the fit to the literature amplitude-phonon data).
The separate and combined effect of each term is illustrated in Fig.~\ref{fig:thzeffect} (c.f. Fig.~4 in the main paper).  In particular, one sees that the additional damping causes the eigenvalues to approach those of the bare modes as $T$ decreases. This reduction in mixing between the (increasingly damped) EOP and the bare phonons can also be understood by considering the mechanical analogy of coupling of oscillators to an overdamped ``bath''.  Also, the pinning potential is seen to play the dominant role for band A in terms the crossing of $\nuLAz$ at $T\approx 120\Kel$ and the initial reduction in bandwidth with decreasing $T$.

%\FloatBarrier 

\subsection{EOP-phonon coupling with both coherent and incoherent contributions}

As mentioned in the main text, the present results and those from the literature \cite{S_tome09} indicate  that only the phase-phonon bands are strongly affected (blue-shifted) by photoexcited free carriers.  The issue was discussed in how to reconcile such a situation within the TDGL model.  As asserted in the main paper, one explanation involves a loss of equivalence in the EOP-phonon coupling $m_n$ for the amplitude and phase channels, due to the coupling losing a strict coherent dependence on the relative phases.  Here we describe how this can be included in the TDGL model quantitatively.

The coupling term $\usr{U}{c}{}$ between the EOP and each bare phonon is replaced by 
\begin{equation}
\usr{U}{c}{} \to -m_n\Delta\cdot\xi_n [ (1-\eta) \cos(\varphi-\chi) + \eta],
\label{eq:incohcoupl}%
\end{equation}
where $0<\eta<1$ represents the loss of coherence.  Physically, this could arise in the photoexcited state due to local fluctuations, whereby the coexistence of the EOP and phonon amplitude still yields a stabilization of the local energy density, only that the strict local phase relation is suppressed.
Using Eq.~\ref{eq:incohcoupl} in Eq.~\ref{eq:potfunc}, one still obtains the same result for the equilibrium given above ($\Delta_0$,$\xi_{0n}$). The Hessian about the equilibrium is however altered, leading to the equations of motion:
\begin{subequations}
\begin{align}
\partial_t^2{\hat\Delta_1}&=-\left[ 2\alpha(\Tc-T) + \frac{m_n^2}{\Ommzsq} \right]\hat\Delta_1
 +m_n\hat\xi_{n1} - \gamma_1\partial_t{\hat\Delta_1}\\
\partial_t^2{\hat\Delta_2}&=-\frac{m_n^2(1-\eta)}{\Ommzsq}\hat\Delta_2
 +m_n(1-\eta)\hat\xi_{n2} - \gamma_2\partial_t{\hat\Delta_2}\\
\partial_t^2{\hat\xi_{n1}}&=m_n\hat\Delta_{1} - \Ommzsq \hat\xi_{n1} \\
\partial_t^2{\hat\xi_{n2}}&=m_n(1-\eta)\hat\Delta_{2} - \Ommzsq \hat\xi_{n2}
\end{align} \label{eq:motionincoh}%
\end{subequations}
Comparison with Eq.s~\ref{eq:motion} shows that the amplitude channel is unaffected (Eq.s~\ref{eq:motionincoh}(a,c)), while the effective coupling for the phase channel is reduced (essentially by a factor $(1-\eta)$, although the first term in Eq.~\ref{eq:motionincoh}(b) corresponds rather to $m_n\to m_n\sqrt{1-\eta}$).  Hence, this modification predicts that the phase-phonons should be shifted toward their bare frequencies upon photoexcitation, as per the dominant trend seen in the data (see Fig.~4 in the main paper).

\section{Comparison of the TDGL model with quantum mechanical models}

It is instructive to compare the qualitative predictions of the TDGL model with those from earlier quantum mechanical treatments (QM) in the literature (based on the Fr{\"o}hlich Hamiltonian).
The extension of the original LRA model \cite{S_lee74} to $N{>}1$ bare phonons \cite{S_rice78} provides a closed-form solution for the $Q=0$ amplitude- and phase-phonon spectra for $T\to 0$.
This yields $N$ renormalized modes for each channel, with $\nuLmzz < \nuLm <\nuLmz$
(as well as the single-particle gap near $2\Delta$ in the conductivity spectrum).
For typical values of the e-ph coupling coefficients, one has: (i) the lowest amplitude mode (``amplitudon'') is most strongly red-shifted; (ii) the lowest phase mode (``phason'') is driven to $\nu=0$ (or close to it, if pinning is included \cite{S_rice78}) leaving only $N{-}1$ renormalized phase-phonons at finite frequency.
This contrasts to the predictions of the TDGL model, where, as detailed above, one has $N$ finite-frequency modes also for the phase channel in the overdamped limit (and an additional mode at $\nu=0$). 
Hence in the nominal QM theory, we would have consider a re-assessment of our assignment for band A (the lowest-frequency phase-phonon), and, for that matter, the basis of the TDGL model.
We note that subsequent extensions of the QM model to include long-range Coulomb terms \cite{S_wong87,S_viro93} predict that the phason resonance can actually break up into both acoustic and optical resonances, the latter of which could provide an alternate assignment for our band A (although the $T$-dependence of the example results in \cite{S_wong87} do not correlate with the frequency/bandwidth behavior of our band here).
Interestingly, if one reduces the phase damping $\gamma_2$ in the TDGL model to the underdamped regime, we find that one obtains a set of phase-phonons qualitatively consistent with the QM model (i.e. with an additional phase-mode at $\nu=0$ and with non-zero damping), although this does not readily aid in interpreting our results here.  Conversely, one could consider that the Fr{\"o}hlich Hamiltonian description does not adequately include the phase damping, and should rather converge to the TDGL result if this were incorporated.  However, in this case, the QM theory would no longer predict the phason at $\nu \sim 0$, where a band with a peak at $\nu\sim 100\ghz$ has indeed been assigned in experimental studies \cite{S_Degi91a}.  Clearly this issue deserves further theoretical investigation, as we assert in the main paper.

\section{Simulations of transient broadening of phonon response around delay zero}

As presented in the main paper, some of the initial spectral features in the reflectivity spectra $\drrel$ (i.e. for $\tau \lesssim 2 \psec$) could not be fitted using the quasi-stationary Drude-Lorentz dispersion model, where a form of transient broadening occurs for the Lorentzian bands.
This phenomena is primarily due to the frequency-mixing which occurs when the THz probe pulse interacts with a system whose polarization response changes on the time scale of the THz cycle period. While these effects have been studied theoretically for Drude response (e.g. \cite{S_nien05}), we did not find any exposition for the perturbation of (narrow) Lorentzian bands, where the temporal response (free-induction decay) is longer and hence temporal-spectral effects can be more severe.
Here we present the simulation results of such OP-TP reflection experiments, for the case of a single Lorentzian band which undergoes a blue-shift in its instantaneous response on a short time scale, to demonstrate how the experimentally observed broadening (as identified in the main paper) manifests.

While the finite-difference time-domain (FDTD) simulations are a proven method to simulate pump-probe experiments with rapidly changing response functions \cite{S_Lars11,S_Thomson13}, these can be time-consuming, especially for batch runs (i.e. vs. pump-probe delay $\tau$).
Instead, here we develop and apply a model for a homogeneously excited thin-film between air (refractive index $n_1$) and substrate ($n_3$) which allows time-integration of only a small number of field/polarization values.
The treatment is based on applying the time-domain boundary conditions for the fields at both interfaces and expanding the field development to first-order in the sample depth ($z\in[0,L]$).
The resulting coupled equations for the reflected (\Ert) and transmitted (\Ett) fields, in terms of the known incident field (\Eit) are,
%\begin{subequations}
\begin{equation}
%\begin{align}
\begin{split}
\frac{\partial \Ert}{\partial t}
&=\frac{\partial \Eit}{\partial t}
+ \frac{c}{n_1 L}(-\Eit + \Ert +\Ett) \\
\frac{\partial \Ett}{\partial t} + \frac{1}{\epsz}\frac{\partial}{\partial t}\Ptt
&=\frac{2n_1 c}{L}(\Eit -\Ert)
-\frac{2n_3 c}{L}\Ett \\
&- \frac{\partial}{\partial t}\left[ \Eit + \frac{1}{\epsz}\Pit \right]
- \frac{\partial}{\partial t}\left[ \Ert + \frac{1}{\epsz}\Prt \right]
\end{split}
%\end{align}
\label{eq:dEdt}
%\end{subequations}
\end{equation}
where each \Pnt{} is the co-integrated time-domain solution to the corresponding polarization equation
\begin{equation}
\frac{\partial^2 \Pnt}{\partial t^2} + \Gamma(t)\frac{\partial \Pnt}{\partial t} 
+\omega_0^2(t)\Pnt = S(t)E_n(t)
\label{eq:dPdt}
\end{equation}
where $\Gamma(t)$, $\omega_0(t)$ and $S(t)$ are the time-dependent damping, resonant frequency and strength of the electronic oscillator (whose temporal development is initiated by a pump pulse centered at $t=-\tau$). These equations are readily generalized to multiple Lorentzian bands and the addition of an instantaneous response due to $\epsbr$.  Their validity was checked by comparison of numerical data for a time-stationary medium with the analytic formula for the reflected/transmitted fields.
To simulate the experimental data, Eq.s~\ref{eq:dEdt}-\ref{eq:dPdt} are integrated for a large set of $\tau$-values.  The resulting raw data (integrated vs. laboratory time $\usr{t}{lab}{}$) are then interpolated onto the experimental time base $t$ (which is skewed relative to $\usr{t}{lab}{}$ due to the use of delay stage in the THz emitter beam path \cite{S_nemec02}), and after time-windowing and Fourier transformation $\drrelx$ is calculated.

\begin{figure}[b]
\includegraphics[width=0.7\columnwidth]{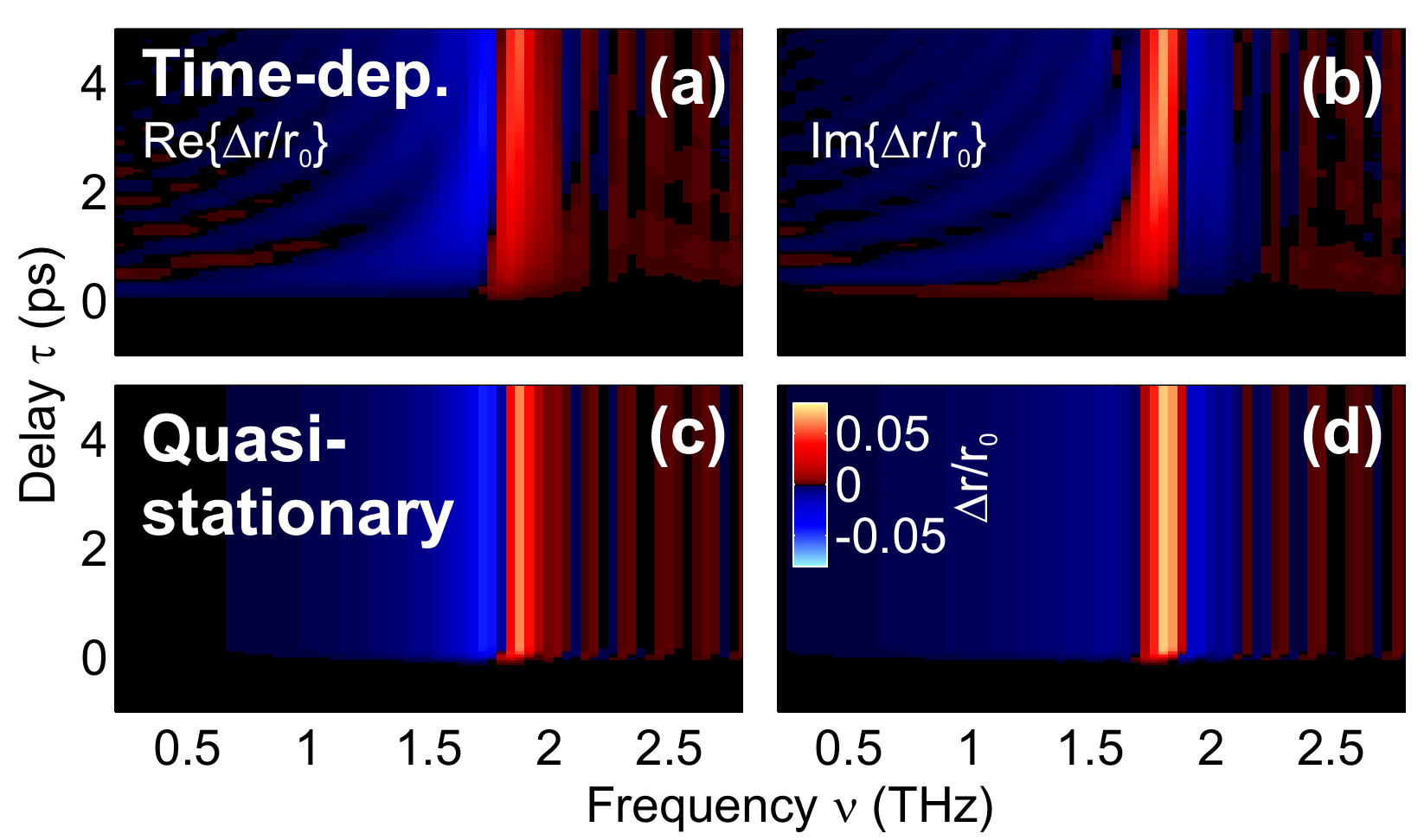}%
\caption{
(a) Simulation of differential reflectivity spectra $\drrelx$ vs. pump-probe delay $\tau$, from integration of time-non-stationary polarization response in a thin layer with homogeneous excitation ($\usr{D}{ex}{}=2\units{\mum}$), for a model system where a ground-state phonon at $\nu=1.75\thz$ undergoes a photo-induced shift to $1.85\thz$ and weakens by 20\% (pump pulse duration $150\fsec$). The substrate refractive index was taken as frequency-independent, $n_3=\sqrt{80}$. (b)~Corresponding idealized quasi-stationary response, to demonstrate how the model band parameters change with delay.  The same temporal time windowing was employed in both cases before Fourier transformation, in order to best represent the experimental data in the main paper (Fig.~2(c-d)).
\label{fig:response}}
\end{figure}

As a representative simulation for comparison with the experimental 2D signals $\drrelx$ (see Fig.~2(c,d) in main paper), we consider the case of a single Lorentzian band, whose frequency and strength are perturbed by excitation with a Gaussian pulse with duration $150\fsec$, as shown in Fig.~\ref{fig:response}(a).  Also shown for comparison is the ``fictitious'' simulated data \drrel{} obtained if one uses a stationary response for each ``frozen'' value of the Lorentzian band parameters (which would be the case for a much slower transition in the parameters).
Clearly one observes the dynamic broadening artifact in the simulated data (especially during the first $\simx 2\psec$), which bears a close resemblance to those in the experimental data (both for the real and imaginary parts of \drrel).
Note that this is an inherent effect in light-matter interaction close to the uncertainty limit (i.e. during rapid evolution compared to the oscillation period, spectral resonances must become broadened), and accentuates that OP-TP transients on a sub-ps time scale cannot be interpreted directly in terms of a quasi-stationary picture.  A more rigorous approach is to consider the response of the perturbed system rather in the dual-frequency domain \cite{S_nemec02}, although this blurs the ability to analyse/interpret the time-evolution of the spectral response (which becomes well-defined again for longer time scales).

%\FloatBarrier 

%\bibliography{kmo_optp_bib1}

%\end{document}

\end{document}